\documentclass[aps,prd,twocolumn,superscriptaddress,preprintnumbers,nofootinbib,showpacs]{revtex4-1}
\usepackage{bm,amsmath,amssymb}
\usepackage{graphicx}
\usepackage{color}
\usepackage{soul}
\usepackage{hyperref}

\setulcolor{blue}
\setstcolor{red}
\sethlcolor{yellow}

\begin{document}

\title{Diagonal and off-diagonal quark number susceptibilities at high temperatures}

\author{H.-T. Ding}
\affiliation{ Key Laboratory of Quark \& Lepton Physics (MOE) and Institute of
Particle Physics, \\ Central China Normal University, Wuhan, 430079, China }
\author{Swagato Mukherjee}
\affiliation{Physics Department, Brookhaven National Laboratory, Upton, NY 11973, USA}
\author{H. Ohno}
\affiliation{Physics Department, Brookhaven National Laboratory, Upton, NY 11973, USA}
\affiliation{ Center for Computational Sciences, University of Tsukuba, Tsukuba,
Ibaraki 305-8577, Japan}
\author{P. Petreczky}
\affiliation{Physics Department, Brookhaven National Laboratory, Upton, NY 11973, USA}
\author{H.-P. Schadler}
\affiliation{Physics Department, Brookhaven National Laboratory, Upton, NY 11973, USA}
\affiliation{Institute of Physics, University of Graz, 8010 Graz, Austria}

\begin{abstract}

We present continuum extrapolated lattice QCD results for up to fourth-order diagonal
and off-diagonal quark number susceptibilities in the high temperature region of
$300-700$ MeV. Lattice QCD calculations are performed using 2+1 flavors of highly
improved staggered quarks with nearly physical quark masses and at four different
lattice spacings. Comparisons of our results with recent weak coupling 
calculations yield reasonably good agreements for the entire temperature range.

\end{abstract}

\maketitle

\section{Introduction}

At high temperatures strongly interacting matter undergoes a deconfinement
transition, where thermodynamics can be described in terms of quarks and gluons
\cite{Petreczky:2012rq,Philipsen:2012nu,Ding:2015ona}. Lattice 
QCD studies  \cite{Petreczky:2012rq,Philipsen:2012nu,Ding:2015ona} as well as
results from heavy-ion collision experiments \cite{Muller:2006ee} suggest that at
least up to temperatures couple of times larger than the transition
temperature quark-gluon plasma (QGP) may behave as a strongly coupled liquid. The
asymptotic freedom of QCD guarantees that at sufficiently high temperatures QGP
becomes weakly coupled and should be described by weak coupling
expansion results. However, nonperturbative effects could still remain important
even at very high temperatures due to the infrared problems arising from the
chromomagnetic sector \cite{Linde:1980ts}. Thus, quantitative validations of weak
coupling QCD calculations against fully nonperturbative lattice QCD results
are necessary to ascertain the temperature range where the strongly coupled QGP
liquid goes over to a weakly coupled quark-gluon gas.

Fluctuations of and correlations among the conserved charges are known to be
sensitive probes of deconfinement and are also suitable for testing the weakly or
strongly coupled nature of QGP.  The study of fluctuations and correlations of
conserved charges on the lattice was initiated some time ago
\cite{Gavai:2002jt,Bernard:2004je,Allton:2003vx,Allton:2005gk}. At low temperature
fluctuations and correlations of conserved charges can be understood in terms of an
uncorrelated hadron gas
\cite{Cheng:2008zh,Bazavov:2012vg,Bazavov:2013dta,Bellwied:2013cta,Borsanyi:2013hza,Bazavov:2014xya,Bazavov:2014yba,Gattringer:2015hra}.
Deconfinement is signaled by a sudden breakdown of such a hadronic description in the
vicinity of the QCD transition temperature \cite{Bazavov:2013dta,Bazavov:2014yba}. 

Fluctuations of and correlations among the conserved charges are defined through the
appropriate derivatives of the pressure, $p$ with respect to the chemical potentials
associated with the corresponding conserved charges. In 2+1 flavor QCD there are
three chemical potentials corresponding to baryon number, electric charge and
strangeness. Since at high temperatures the relevant degrees of freedom are quarks
and gluons, it is natural to use the flavor chemical potentials corresponding to the
$u$ (up), $d$ (down) and $s$ (strange) quark numbers instead of the three conserved
charge chemical potentials. In the flavor basis the fluctuations and correlations of
charges get mapped into the diagonal ($\chi_n^q$) and the off-diagonal
($\chi_{nm}^{qq^\prime}$) quark number susceptibilities
\begin{equation}
\chi_{n}^{q}= \frac{\partial^{n}\left(p/T^4\right)} {\partial\left(\mu_q/T \right)^n},
~\mathrm{and}~
\chi_{nm}^{qq^\prime}= \frac{\partial^{n+m}\left(p/T^4\right)}
{\partial\left(\mu_q/T\right)^n \left(\mu_{q^\prime}/T\right)^m} .
\end{equation}
Here, $\mu_q$ and $\mu_{q^\prime}$ are the chemical potentials corresponding to quark
flavor $q$ and $q^\prime$, with $q,~q^\prime=u,d,s$, and $n,m$ denote the number of
derivatives taken with respect to the quark flavors.

At sufficiently high temperatures the diagonal and off-diagonal quark number
susceptibilities should be described by weak coupling expansion results.
Second-order quark number susceptibilities have been studied in weak coupling
expansion for some time
\cite{Blaizot:2001vr,Vuorinen:2002ue,Rebhan:2003fj,Hietanen:2008xb}. Recently also
weak coupling results for the fourth-order fluctuations and correlations have been
presented
\cite{Haque:2010rb,Andersen:2012wr,Mogliacci:2013mca,Haque:2013qta,Haque:2013sja,Haque:2014rua}.
The validity of these weak coupling expansion results for the second-order diagonal
susceptibilities $\chi_2^q$ has been tested against the accurate continuum
extrapolated lattice QCD results in Ref. \cite{Bazavov:2013uja}. Here we extend this
previous study to perform accurate continuum extrapolated lattice QCD calculations of
fourth-order diagonal quark number susceptibility $\chi_4^q$ and subsequently compare them
with weak coupling perturbative QCD results to validate the weak coupling regime of
QGP. Diagrammatically, the off-diagonal quark number susceptibilities only consist
of quark-line disconnected diagrams and corrections to the tree level generically start
at higher orders of the coupling. Thus, these off-diagonal susceptibilities provide
more stringent tests of the weak coupling regime of QGP. Our present study also
includes lattice calculations and their comparisons with weak coupling results of up
to fourth-order off-diagonal quark number susceptibilities. 

The rest of the paper is organized as follows. In Sec. II we describe the details
of our lattice QCD calculations. Sec. III is dedicated to the discussion of fourth
order diagonal quark number susceptibilities, including the details of continuum
extrapolations and comparison with weak coupling approaches. In Sec. IV we present
our results for the off-diagonal quark number susceptibilities.  Finally, Sec. V
contains our conclusions.

\section{Details of the lattice simulations}

We performed lattice calculations in 2+1 flavor QCD using the highly improved
staggered quark (HISQ) action \cite{Follana:2006rc}. Lattice sizes were chosen to be
$N_{\sigma}^3 \times N_{\tau}$ with $N_{\tau}=6,~8,~10$ and $12$ and a fixed aspect
ratio of $N_{\sigma}/N_{\tau}=4$. The gauge configurations used in this study were
generated by the HotQCD Collaboration using the physical value of the strange quark
mass $m_s$ and degenerate up and down quark masses $m_u=m_d=m_l=m_s/20$
\cite{Bazavov:2011nk,Bazavov:2014pvz}. The latter corresponds to a pion mass of
$161$~MeV in the continuum limit \cite{Bazavov:2011nk}. For each value of $N_{\tau}$
the temperature was varied by varying the lattice spacing $a$ or equivalently the
lattice gauge coupling $\beta=10/g^2$. The lattice spacing has been fixed using the
$r_1$ scale defined in terms of the static quark potential
\begin{equation}
	\left.r^2 \frac{dV}{dr}\right|_{r=r_1}=1.
\end{equation}
We used the parametrization of $a/r_1$ from Ref.~\cite{Bazavov:2014pvz}.  As in
Ref.~\cite{Bazavov:2011nk} we use $r_1=0.3106$~fm to convert to physical units. To
extend the temperature coverage in our calculations additional gauge configurations
have been generated for $\beta=10/g^2=8.0,~8.2$ and $8.4$ on $32^3 \times 8$, $40^3
\times 10$ and $48^3 \times 12$ lattices.  As in
Refs.~\cite{Bazavov:2011nk,Bazavov:2014pvz} we use the Rational Hybrid Monte Carlo
algorithm \cite{Clark:2004cp}.  About $2000-4000$ gauge configurations separated by
ten molecular dynamic trajectories of unit length have been generated for each value
of the gauge coupling $\beta$.  The complete list of the gauge configurations used in
this study can be found in the Appendix.

The quark number susceptibilities can be expressed in terms of the quark matrix and its inverse,
and the corresponding formulas were given in Refs.
\cite{Allton:2002zi,Allton:2003vx,Allton:2005gk}.  The necessary operators are
evaluated using the random noise method with unbiased estimators (see
Ref.~\cite{Allton:2002zi} for details).  We used between $150$ to $250$ random source
vectors to evaluate the needed operators depending on the lattice volume and the
value of $\beta$.  We have found that this number of source vectors is sufficient to
ensure that the noise due to gauge field fluctuations is larger than the noise from
the stochastic estimators. In a few cases, some operators have been estimated with many
more random source vectors to ensure that there are no additional systematic errors
due to the limited number of source vectors.  In the Appendix we give a detailed
account for the number of random source vectors used in our study.
The inversion of the quark matrix was performed in double precision, and
the squared residual of the inversion was less than $10^{-19}$.

\begin{figure}
\includegraphics[width=8cm]{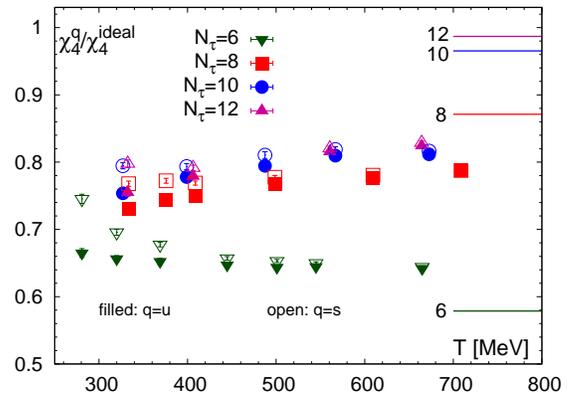}
\caption{The light ($q=u$) and strange ($q=s$) fourth-order quark number
susceptibilities shown as filled and open symbols, respectively, for different values
of the lattice temporal extent $N_{\tau}$ and normalized by the results for the
continuum, massless ideal gas limit $\chi_4^{\rm ideal}=6/\pi^2$. The horizontal
lines denote the lattice ideal gas results for different values of $N_\tau$.
}
\label{fig:chi4}
\end{figure}

\section{Diagonal quark number susceptibilities}

\begin{figure*}
\includegraphics[width=5.5cm]{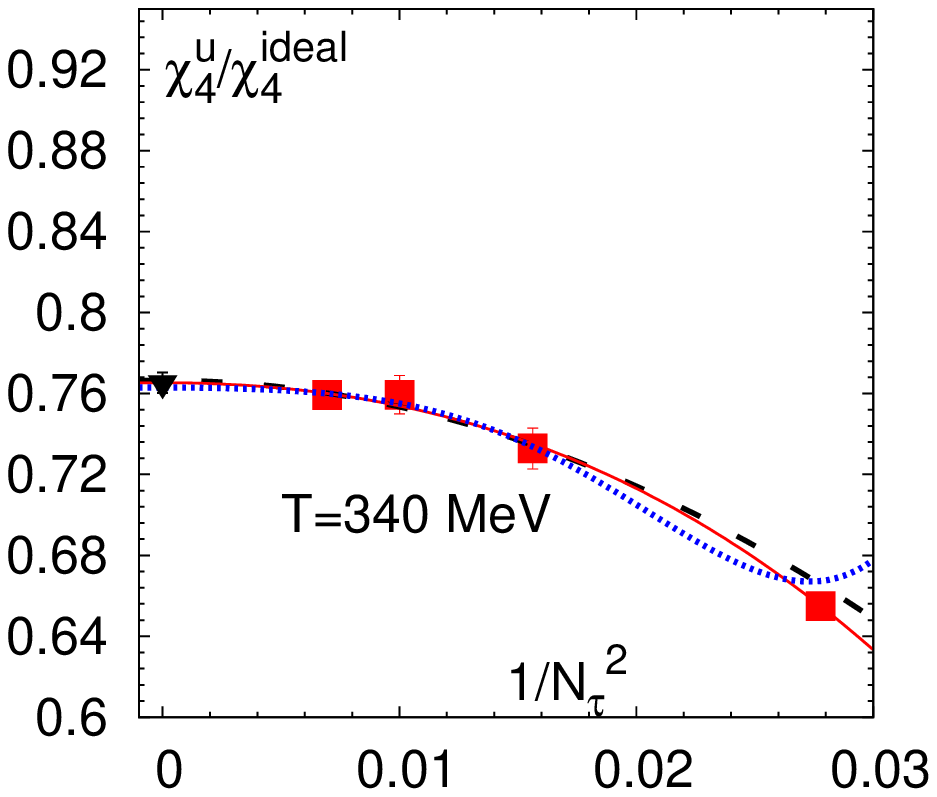}
\includegraphics[width=5.5cm]{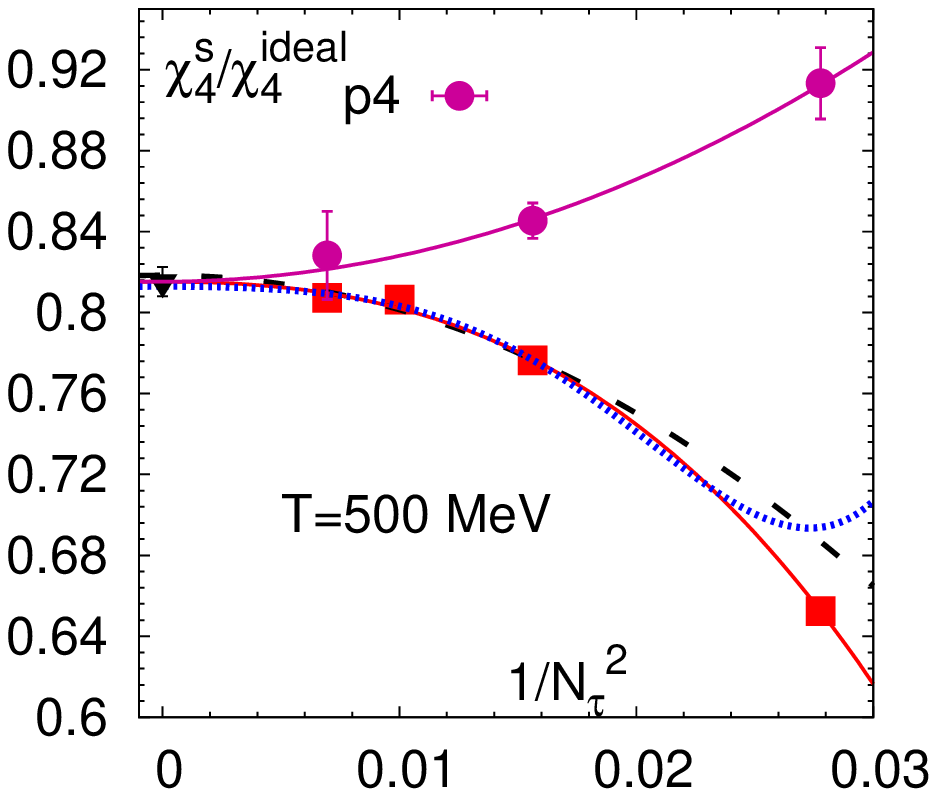}
\includegraphics[width=5.5cm]{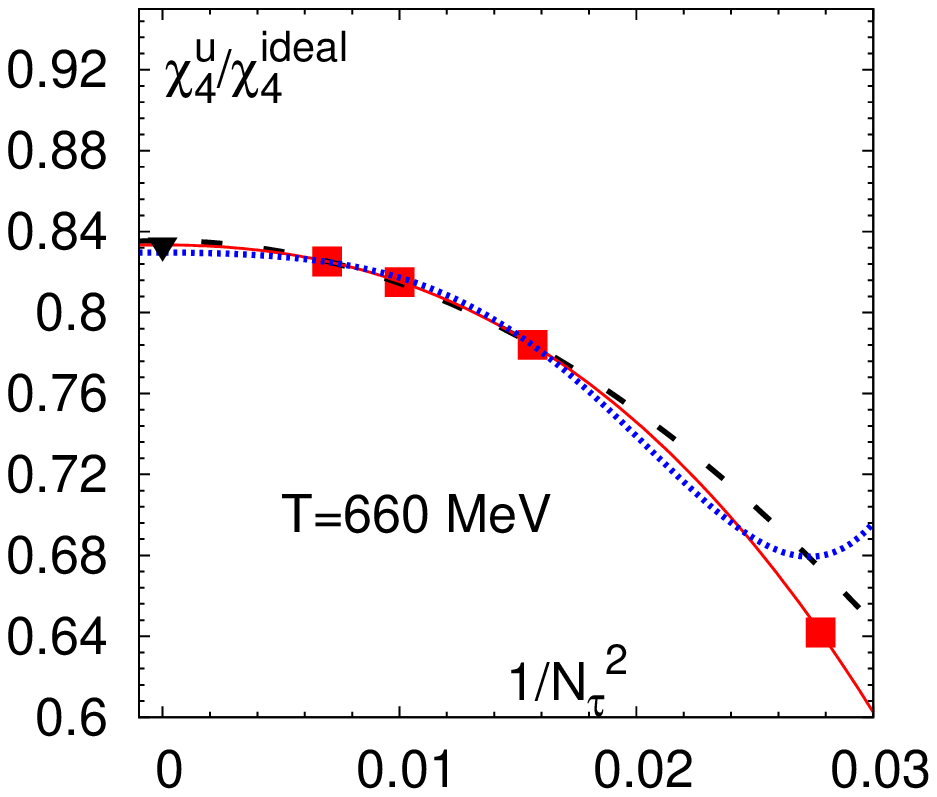}
\caption{Continuum extrapolations of $\chi_4^q$ at three representative temperatures.
In the middle panel, we show extrapolation for $\chi_4^s$ and compare our results with
the $N_{\tau}$ dependence of $\chi_4^s$ obtained with p4 action
\cite{Bazavov:2013uja}. The solid and dashed lines correspond to continuum
extrapolations using Eq.  (\ref{extra}), while the dotted lines correspond to
extrapolations performed using Eq. (\ref{extra_free}); see the main text. The filled
triangle corresponds to the continuum value. All the results have been normalized by
$\chi_4^{\rm ideal}=6/\pi^2$. 
}
\label{fig:extra}
\end{figure*}

In this section we will discuss our results on diagonal quark number
susceptibilities. The second-order light and strange quark number susceptibilities
have been discussed in Ref.~\cite{Bazavov:2013uja} in great detail using HISQ action,
including the continuum extrapolation. Our results for the second-order quark number
susceptibilities agree with the findings of Ref.~\cite{Bazavov:2013uja}, and therefore
we will not discuss these further. In Fig.~\ref{fig:chi4} we show our results for the
light and strange fourth-order quark number susceptibilities $\chi_4^q$ ($q=u,s$) for
$N_{\tau}=6,~8,~10$ and $12$. The cutoff dependence of $\chi_4^q$ is similar to
that of $\chi_2^{u,s}$; the continuum is approached from below as suggested by the
cutoff effects in the lattice free theory. Similar to the case of $\chi_2^q$
\cite{Bazavov:2013uja}, the size of the cutoff effects is smaller than in the free
theory,  cf. Fig.~\ref{fig:chi4}. The quark mass dependence is, however,
opposite to the case of $\chi_2^q$ for $T< 500$~MeV. Namely, the fourth-order light
quark number susceptibilities are below the strange quark number susceptibilities,
while for the second-order susceptibilities $\chi_2^s<\chi_2^u$
\cite{Bazavov:2013uja}. 
For $T > 500$ MeV the quark mass effects in the fourth-order quark number
susceptibilities are smaller than the statistical uncertainty.

To perform the continuum extrapolation, we first interpolate the lattice results as a
function of the temperature using smoothing splines. The spline interpolation is
performed using the R-package \cite{Rpackage}.  The errors of the spline are
estimated using the bootstrap method. At selected values of the temperature in the
interval of $T=340$~MeV to $T=660$~MeV, separated by $20$~MeV, we performed continuum
extrapolations using the interpolated values of $\chi_4^q$ and the corresponding
bootstrap errors. Illustrative results for the continuum extrapolations for
$\chi_4^q$ are shown in Fig.~\ref{fig:extra}. For each temperature we performed the
continuum extrapolation using the form motivated by the lattice free theory
\cite{Hegde:2008nx},
\begin{equation}
\chi_4^q(N_{\tau})=a+b/N_{\tau}^4+c/N_{\tau}^6.
\label{extra}
\end{equation}
We also performed extrapolations using only the data for $N_{\tau} \ge 8$ and setting
$c=0$. These extrapolations typically give larger continuum values but remain
compatible within errors with the original extrapolations. These fits are shown as
solid and dashed lines, respectively, in Fig.~\ref{fig:extra}.  We do not find
evidence for a significant $1/N_{\tau}^2$ term in the cutoff dependence for
$\chi_4^q$. This is evident from the $N_{\tau}$ dependence of our results shown in
Fig.~\ref{fig:extra}. A similar situation was observed for $\chi_2^q$
\cite{Bazavov:2013uja}. Since the cutoff dependence of $\chi_4^q$ qualitatively
follows the lattice free theory expectation we also performed continuum extrapolation
using the form
\begin{equation}
\chi_4^q(N_{\tau}) = e +f \cdot (\chi_4^{q,{\rm lat-ideal}}(N_{\tau})-\chi_4^{\rm
ideal}),
\label{extra_free}
\end{equation}
where $\chi_4^{q,{\rm lat-ideal}}(N_{\tau})$ is the lattice free theory result and
$\chi_4^{{\rm ideal}}=6/\pi^2$ is the continuum free theory result for the
fourth-order quark number susceptibility.  This simple two-parameter fit describes
the data well.  The results of these fits are also shown in Fig.~\ref{fig:extra} as
dotted lines. For the coefficient $f$ we get values ranging from about $0.30$ at the
lowest temperature to about $0.43$ at the highest temperature;  i.e., cutoff
effects are only $30\%-40\%$ of those in the free theory. The central
values of the extrapolations obtained from Eq. (\ref{extra_free}) are typically
smaller than the ones obtained from the three-parameter fit with Eq. (\ref{extra}),
but still compatible within errors. In other words, there are no statistically
significant deviations between the different continuum extrapolations. Therefore, the
extrapolated values obtained from three-parameter fits with Eq. (\ref{extra}) will be
used as our continuum results. Since the different continuum extrapolations give
a difference of the order of the statistical errors, we conservatively estimate the
total error on our continuum results for $\chi_4^q$ as twice the statistical error
obtained from our three-parameter fits.

\begin{figure}
\includegraphics[width=8cm]{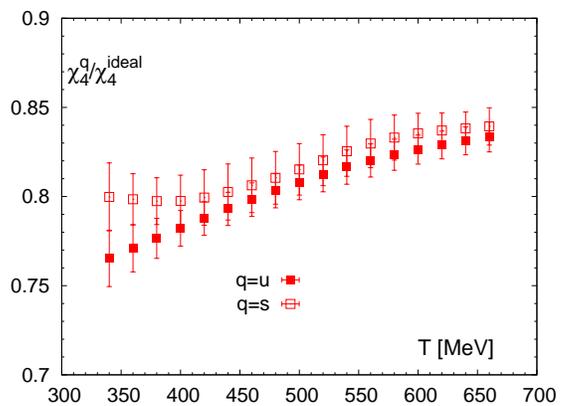}
\caption{The continuum extrapolated results for $\chi_4^u$ and $\chi_4^s$ normalized
by the corresponding massless ideal gas value $\chi_4^{\rm ideal}=6/\pi^2$.}
\label{fig:chi4_cont}
\end{figure}

In Fig.~\ref{fig:extra} we also show the lattice results for $\chi_4^s$ for
$T=500$~MeV obtained with p4 action for $N_{\tau}=6,~8$ and $12$
\cite{Bazavov:2013uja}. As discussed in Ref.~\cite{Bazavov:2013uja}, the continuum
limit for p4 action is approached from abovei, and the cutoff dependence in this case
does not follow the free theory expectation; the leading-order cutoff dependence goes
like $1/N_{\tau}^2$. As was also discussed therei, the quality of the p4 data does not
allow a reliable continuum extrapolation for $\chi_4^s$. However, in the continuum
limit, the results obtained with HISQ action and p4 action should of course agree.
Therefore, we fit the cutoff dependence of the p4 data using the form
$a+g/N_{\tau}^2+h/N_{\tau}^4$ and requiring that in the continuum limit it agrees
with the HISQ result. This fit gives $\chi^2/{\rm d.o.f.} \simeq 0.1$. Therefore, we
can at least say that the p4 data are consistent with the continuum extrapolations
performed with HISQ action. The continuum extrapolated results for $\chi_4^u$ and
$\chi_4^s$ are shown in Fig.~\ref{fig:chi4_cont} as a function of the temperature. We
can see that $\chi_4^s>\chi_4^u$ but the difference between $\chi_4^s$ and $\chi_4^u$
becomes smaller with increasing temperature. The continuum extrapolated results for
$\chi_4^q$ are provided in Table \ref{tab:cont}.

\begin{figure}
\includegraphics[width=8cm]{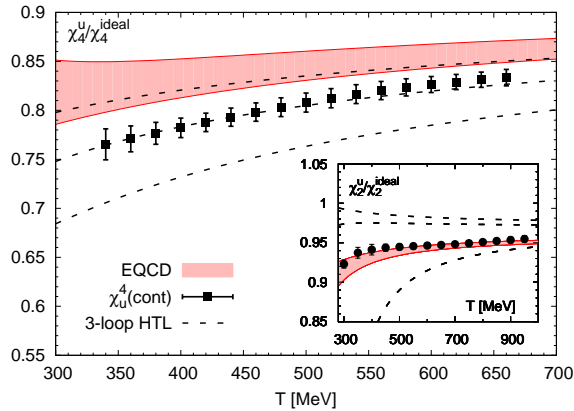} 
\caption{The continuum extrapolated result for $\chi_4^u$ compared to perturbative
EQCD calculations shown as the shaded band. The width of the band corresponds to the
variation of the renormalization scale from $\pi T$ to $4 \pi T$. The dashed lines
correspond to the three-loop HTL calculations evaluated for the renormalization scale
$\Lambda=4 \pi T$, $2 \pi T$ and $\pi T$ (from top to bottom). All results have been
normalized by the corresponding massless ideal gas results of $\chi_4^{\rm
ideal}=6/\pi^2$. 
The inset shows the comparison of the lattice and weak coupling calculations
for $\chi_2^u$ \cite{Bazavov:2013uja} normalized by $\chi_2^{\rm ideal}=1$.
} 
\label{fig:chi4_pert} 
\end{figure}

The deviation of $\chi_4^q$ from the ideal gas limit at the highest temperature
studied by us is 15\%.  This should be compared to the continuum result for
$\chi_2^q$ which is only $6\%$ below the ideal gas limit at a similar temperature
\cite{Bazavov:2013uja}. It is interesting to see whether these deviations from the
ideal gas limit can be understood in terms of weak coupling expansions. The fourth
order quark number susceptibilities have been calculated in three-loop hard thermal loop
(HTL) perturbation theory recently in Ref.~\cite{Haque:2014rua} and in perturbative
calculations using dimensionally reduced electrostatic QCD (EQCD)
\cite{Mogliacci:2013mca}. These weak coupling results are compared with our continuum
results for $\chi_4^u$ in Fig.~\ref{fig:chi4_pert}. We show the scale uncertainty of
the perturbative results by varying the renormalization scale $\Lambda$ from $\pi T$
to $4 \pi T$. The comparison of the continuum lattice results for $\chi_2^u$ 
\cite{Bazavov:2013uja} with weak coupling calculations is also shown in Fig.~\ref{fig:chi4_pert} as an inset.
The EQCD band is above the continuum extrapolated lattice data for
$\chi_4^u$, while these calculations give results for $\chi_2^u$ that agree with the
lattice data. So there remains some tension between the EQCD calculations and the
lattice results on second and fourth-order quark number susceptibilities. The three-loop
HTL perturbative result agrees very well with our lattice data for the
renormalization scale $\Lambda=2 \pi T$. This scale choice, however, overpredicts
the value of $\chi_2^u$. Although the scale uncertainty in the three-loop HTL calculation
of $\chi_2^u$ is rather large, it should be noted that, once the renormalization scale
$\Lambda$ is fixed by comparing one observable with the corresponding lattice QCD
result all the other quantities are completely parameter-free predictions of HTL
calculations without any uncertainty. Unfortunately, the lattice QCD results for
$\chi_2^q$ and $\chi_4^q$ cannot be simultaneously reproduced by the three-loop HTL
calculations with a single renormalization scale for the coupling. Overall the three-loop
HTL perturbative results agree with the lattice data within a somewhat large scale
uncertainty. 

\section{Off-diagonal quark number susceptibilities}

\begin{figure*}
\includegraphics[width=7cm]{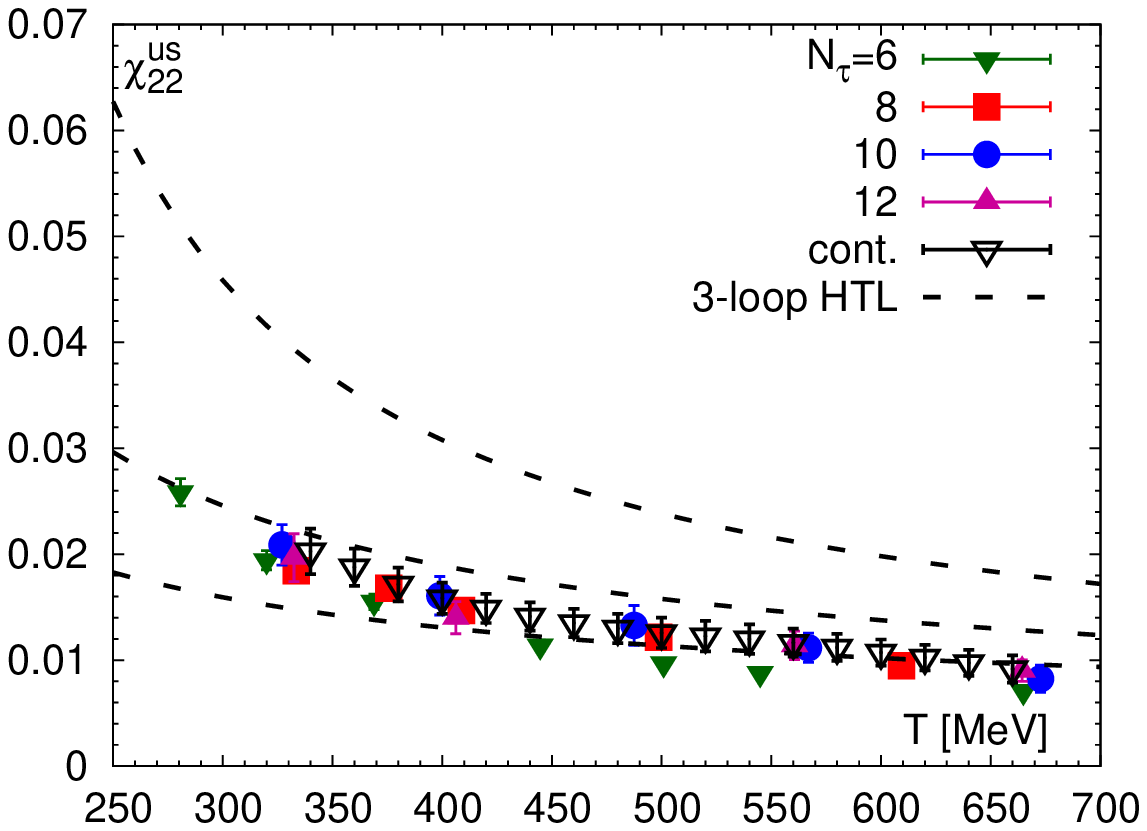}
\includegraphics[width=7cm]{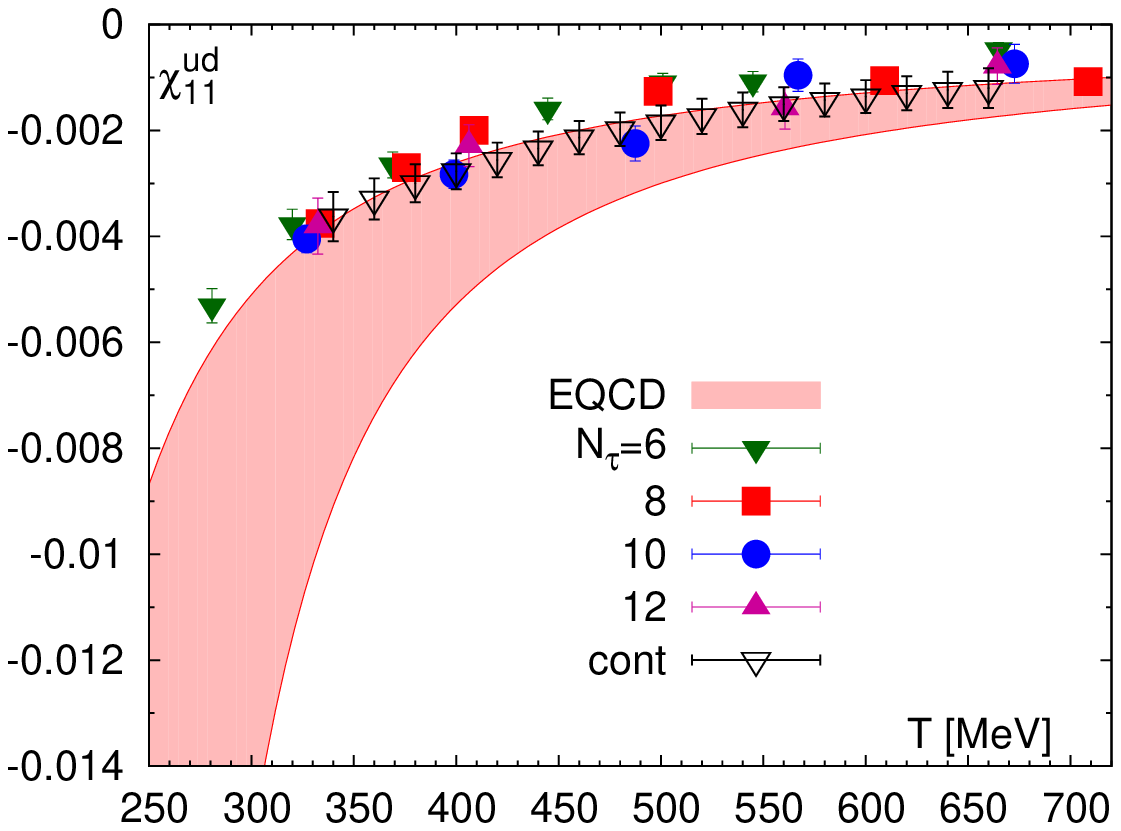}
\caption{The fourth-order off-diagonal susceptibility $\chi_{22}^{us}$ (left) and the second
order off-diagonal quark number susceptibility $\chi_{11}^{ud}$ (right) calculated
for different values of $N_{\tau}$. The lattice results are compared to EQCD
calculations \cite{Hietanen:2008xb} shown as the shaded band corresponding
to renormalization scale variation from $\pi T$ to $4 \pi T$ and to three-loop HTL
perturbation theory \cite{Haque:2014rua} shown with the dashed lines for
renormalization scale $\Lambda=\pi T$, $2 \pi T$ and $4 \pi T$ (from top to bottom).}
\label{fig:offdiag}
\end{figure*}

In this section we show our results on off-diagonal quark number susceptibilities and
compare them with weak coupling calculations. The off-diagonal susceptibilities
vanish in the infinite temperature limit, i.e., for the ideal quark gas. In
fact, using weak coupling calculations, one can show that they are related to the coupling
of quarks to soft (static) gluons. In the language of EQCD, one may say that they do
not receive contribution from the scale $2\pi T$. In particular, the leading-order
contribution to $\chi_{22}^{qq^\prime}$ comes from the term proportional to $T m_D^3$
(the so-called plasmon term) in the expression of the pressure, with $m_D$ being the
leading-order Debye mass. Thus, the leading-order contribution to
$\chi_{22}^{qq^\prime}$ is of order $g^3 \sim \alpha_s^{3/2}$. The leading-order
contribution to $\chi_{11}^{qq^\prime}$ is order $g^6 \sim \alpha_s^3$ and is
sensitive to the static chromomagnetic sector. Therefore, it is nonperturbative and
can only be calculated using lattice simulations of EQCD \cite{Hietanen:2008xb}.
Since the off-diagonal quark number susceptibilities at high temperature are mostly
sensitive to soft gluon fields, the cutoff ($N_{\tau}$) dependence of these quantities
in lattice calculations using HISQ is expected to be small at high temperatures. Our
lattice QCD results for $\chi_{11}^{ud}$ and $\chi_{22}^{us}$ are shown in
Fig.~\ref{fig:offdiag}.  As expected, the apparent cutoff dependence of the
off-diagonal quark susceptibilities is considerably smaller than for the diagonal
ones. The off-diagonal susceptibilities are small, which is in qualitative agreement
with weak coupling expectations discussed above.  Since the cutoff dependence of the
off-diagonal susceptibilities is quite small, it is possible to perform a comparison
with weak coupling calculations using the lattice data at fixed $N_{\tau}$.
Nevertheless, we performed continuum extrapolations also for off-diagonal quark
number susceptibilities.  The procedure of continuum extrapolation here is similar to
that for diagonal quark number susceptibilities. We first perform an interpolation in
the temperature and then continuum extrapolations at the same temperature values as
before separated by $20$ MeV.  We performed continuum extrapolations fitting the
$N_{\tau}$ dependence of the lattice data with $a+b/N_{\tau}^2+c/N_{\tau}^4$. We also
performed fits using only $N_{\tau}\ge 8$ data and setting $c=0$. Both fits give
result that agree well within the errors.  In the case of $\chi_{22}^{us}$ it was
also possible to set $b=0$, i.e.,  perform fits to a constant for $N_{\tau} \ge
8$. These fits give the result with the smallest errors but are consistent with the above fits.
In the case of $\chi_{11}^{ud}$, we performed fits with $c=0$ to obtain the
continuum
estimate. In addition averaging over $N_{\tau}=10$ and $N_{\tau}=12$ data
for $\chi_{11}^{ud}$
gave results which are consistent with the above continuum estimate.
We used the two-parameter fits and the corresponding errors for our final
continuum
estimates for $\chi_{22}^{ud}$ and $\chi_{11}^{ud}$.

These continuum estimates are also shown in Fig. \ref{fig:offdiag}. The numerical values of the
continuum extrapolated off-diagonal susceptibilities are also presented in Table
\ref{tab:cont}. We also calculated $\chi_{11}^{us}$ and $\chi_{22}^{ud}$ and
found that within the estimated errors they agree with $\chi_{11}^{ud}$ and
$\chi_{22}^{us}$.

Now let us compare our lattice results with weak coupling calculations in more
details. In Fig.~\ref{fig:offdiag} we also show the results from three-loop HTL
perturbation theory for $\chi_{22}^{us}$ using three choices of the renormalization
scale $\Lambda= \pi T, ~2 \pi T$ and $4 \pi T$ \cite{Haque:2014rua}. The scale choice
$\Lambda = 4 \pi T$ works best for the higher temperature, while for a lower
temperature, the choice $\Lambda =2 \pi T$ seems to be better. Overall it is fair to
say thati, within the uncertainties, the lattice and the three-loop HTL perturbation theory
results agree. 
We note that the three-loop HTL calculation for quark number susceptibilities has
been performed in the limit of vanishing quark masses. As stated above the quark
mass effects are negligable for $\chi_{22}^{qq'}$, and therefore comparison
of three-loop HTL results and the lattice results for $\chi_{22}^{us}$ is justified.
We compare our results for $\chi_{11}^{ud}$ with EQCD calculations of
Ref.~\cite{Hietanen:2008xb}. In EQCD $\chi_{11}^{ud}$ can be written as 
\begin{equation}
\chi_{11}^{ud}=\frac{g_E^6}{\pi^3} \left( \frac{5}{288 \pi^4} \ln (4 y)+
\frac{5}{48 \pi^4} \beta_M + \delta \chi_3(y)\right) \;, 
\end{equation}
where $y=m_E^2/g_E^4$, $g_E$ and $m_E$ are the gauge coupling and the mass parameter
of EQCD which can be calculated at any order in perturbation theory
\cite{Braaten:1995jr}.  The constant $\beta_M$ is a nonperturbative constant that
determines the leading-order contribution to $\chi_{11}^{ud}$. We use the value
$\beta_M=0.1$ determined in Ref.~\cite{Hietanen:2008xb}. The function $\delta
\chi_3(y)$ parameterizes higher order corrections and was calculated in
Ref.~\cite{Hietanen:2008xb}.  Using this as well as the next-to-leading-order expression for $y$ from
Ref.~\cite{Kajantie:1997tt} together with the next-to-next-leading-order result for $g_E$
\cite{Laine:2005ai}, we calculate the EQCD result for $\chi_{11}^{ud}$.  This result
is shown in Fig.~\ref{fig:offdiag} as a band.  The width of the band for
$\chi_{11}^{ud}$ shown in Fig.~\ref{fig:offdiag} corresponds to the error on $\delta
\chi_3$ and the scale variation between $\Lambda=\pi T$ and $\Lambda=4 \pi T$
combined. We see that there is a fair agreement between the lattice results and the
EQCD calculation for $\chi_{11}^{ud}$. 

We also calculated the three other fourth-order off-diagonal quark number
susceptibilities, $\chi_{31}^{ud}$,  $\chi_{31}^{us}$ and $\chi_{13}^{us}$.
Unfortunately, these quantities are very noisy, but within large errors appear to be
both temperature independent as well as $N_{\tau}$ independent. If we fit all the
available lattice data in the temperature range $T=280-710$ MeV with a constant, we
obtain the following estimates:
\begin{eqnarray}
\chi_{31}^{ud} &=& 0.00020  \pm  0.00015 \;, \\
\chi_{31}^{us} &=& 0.00071  \pm  0.00030 \;,\\
\chi_{13}^{us} &=& 0.00060  \pm  0.00030 \;.
\end{eqnarray}
We see that these off-diagonal susceptibilities are similar in magnitude to
$\chi_{11}^{ud}$ which may be expected in the weak coupling calculations.

\section{Conclusions}

In this paper we have extended the study of lattice QCD calculations quark number
susceptibilities at high temperatures using the HISQ action to include fourth-order
diagonal as well as off-diagonal quark number susceptibilities. In our calculations we use
lattices with temporal extent $N_{\tau}=6,~8,~10$ and $12$, and we cover 
temperatures ranging from $300$ up to $700$~MeV. We have obtained sufficiently
accurate reliable continuum extrapolated lattice QCD results at high enough
temperatures to perform meaningful comparisons with weak coupling QCD
calculations.  We find that the cutoff dependence of the fourth-order diagonal quark
number susceptibilities $\chi_4^q$ is similar to the cutoff dependence of the second
order quark number susceptibility $\chi_2^q$ \cite{Bazavov:2013uja}, and continuum
extrapolations can be performed in a similar manner. The systematic errors of the
continuum extrapolations are quite small, smaller than or equal to the statistical errors.
We compared our continuum extrapolated results for $\chi_4^u$ to three-loop HTL
perturbation theory as well as to EQCD results. The result of three-loop HTL perturbation
theory agrees very well with our continuum extrapolated result for $\chi_4^u$.
However, three-loop HTL calculations cannot simultaneously reproduce the lattice QCD
results for $\chi_2^q$ and $\chi_4^q$ with the same choice of the renormalization
scale for the coupling. The EQCD result of Ref.~\cite{Mogliacci:2013mca} for
$\chi_4^u$ is somewhat above our continuum lattice result, while it agrees with the
continuum lattice results for $\chi_2^u$ \cite{Bazavov:2013uja}.

In the case of the off-diagonal susceptibilities, we also find decent agreements
between the lattice and the weak coupling results for $\chi_{11}^{ud}$ and
$\chi_{22}^{ud}$. From our analysis it is clear that lattice results provide
stringent tests for different weak coupling approaches at high temperatures. The
agreement between the lattice and the weak coupling results suggests that for
$T>300$~MeV quark degrees of freedom are weakly coupled.

While this paper was being finalizedi, Ref. \cite{Bellwied:2015lba} appeared. Reference
\cite{Bellwied:2015lba} performed similar lattice QCD calculations with a different
fermionic discretization scheme and compared up to fourth-order diagonal and
off-diagonal susceptibilities with the weak coupling results. The lattice results and
hence the conclusions of Ref. \cite{Bellwied:2015lba} is very similar to our present
study.

\acknowledgments{
This work was supported by U.S. Department of Energy under Contract No. DE-SC0012704.
H.-P. Schadler was funded by the FWF DK W1203, ``Hadrons in Vacuum, Nuclei and Stars".
The authors thank F.~Karsch and P.~Steinbrecher for interesting discussions.
The numerical computations have been carried out on the clusters of USQCD
collaboration, on the BlueGene supercomputers at the New York Center for
Computational Sciences (NYCCS) and on Vienna Scientific Cluster (VSC).  The calculations
have been performed using the publicly available MILC code. We thank S.~Mogliacci and
N.~Haque for sending the numerical values of their calculations.
}

\begin{table}
\caption{The continuum estimates for the diagonal and off-diagonal
quark number susceptibilities.}
\begin{tabular}{ccccc}
\hline
$T$ [MeV] & $\chi_4^u$ &  $\chi_4^s$ &  $\chi_{11}^{ud}\cdot 10^3$ &  $\chi_{22}^{ud}$ \\
\hline
340 & 0.4653(48)        & 0.4862(58) &  -3.62(46)       & 0.0203(21) \\
360 & 0.4687(41)        & 0.4854(43) &  -3.29(39)       & 0.0188(18) \\
380 & 0.4721(34)        & 0.4848(40) &  -2.99(36)       & 0.0171(16) \\
400 & 0.4755(30)        & 0.4849(44) &  -2.77(33)       & 0.0158(15) \\
420 & 0.4789(29)        & 0.4860(47) &  -2.56(33)       & 0.0149(14) \\
440 & 0.4821(28)        & 0.4879(47) &  -2.34(32)       & 0.0141(13) \\
460 & 0.4853(29)        & 0.4902(47) &  -2.13(31)       & 0.0135(13) \\
480 & 0.4883(29)        & 0.4927(45) &  -1.98(32)       & 0.0130(14) \\
500 & 0.4912(30)        & 0.4956(44) &  -1.85(33)       & 0.0126(14) \\
520 & 0.4939(30)        & 0.4987(43) &  -1.73(33)       & 0.0123(14) \\
540 & 0.4964(29)        & 0.5018(43) &  -1.61(33)       & 0.0120(14) \\
560 & 0.4986(28)        & 0.5044(41) &  -1.50(32)       & 0.0117(13) \\
580 & 0.5006(27)        & 0.5065(38) &  -1.42(31)       & 0.0112(17) \\
600 & 0.5023(25)        & 0.5079(34) &  -1.36(31)       & 0.0107(12) \\
620 & 0.5039(24)        & 0.5089(30) &  -1.30(32)       & 0.0102(12) \\
640 & 0.5053(24)        & 0.5096(28) &  -1.24(34)       & 0.0097(12) \\
660 & 0.5067(26)        & 0.5103(32) &  -1.20(38)       & 0.0091(13) \\
\hline
\end{tabular}
\label{tab:cont}
\end{table}

\appendix* 
\section{Details of lattice calculations}

In this Appendix we present some details of our calculations.  In Tables
\ref{ensemblesnt6}, \ref{ensemblesnt8}, \ref{ensemblesnt10} and \ref{ensemblesnt12} we
show the gauge coupling $\beta=10/g^2$ and the corresponding temperatures for
different lattice sizes. The number of gauge configurations used in the analysis is
shown as the last column of the tables.  We estimate the fermion operators involving
light ($u$ or $d$) quarks using $N_{sv,l}$ random source vectors, while for the
operators involving strange quarks we use $N_{sv,s}$ source vectors. We tried to
improved the stochastic estimates by using additional source vectors for 
all operators that only require two inversions. In most cases these operators
turn out to be the noisiest but because of fewer inversions are cheaper to calculate.
The number of additional source vectors is denoted by $N_{sv,l}^{imp}$ and
$N_{sv,s}^{imp}$ for light and strange quark operators, respectively. For some
ensembles we used many more source vectors to check for systematic errors.

\begin{table}[!h]
\caption{ 
Parameters of the calculations for $24^3 \times 6$ lattices}
\label{ensemblesnt6}
\begin{tabular}{ccccccc}
		\hline 
		$\beta$ & $T$ (MeV) & $N_{sv,l}$ & $N_{sv,l}^{imp}$ & $N_{sv,s}$ & $N_{sv,s}^{imp}$& \# configurations \\
		\hline
	        6.664 &	281.1	& 150	& 0	& 150	& 0	& 2990 \\
					6.800 &	320.5	& 150	& 0	& 150	& 0	& 3000 \\
					6.950 &	369.8	& 150	& 0	& 150	& 0	& 3000 \\
	        7.150 &	445.9	& 150	& 0	& 150	& 0	& 2540 \\					
	        7.280 &	502.5	& 150	& 0	& 150	& 0	& 2720 \\
	        7.373 &	546.7	& 150	& 0	& 150	& 0	& 2310 \\
	        7.596 &	667.2	& 150	& 0	& 150	& 0	& 3000 \\
                 \hline
	\end{tabular}
\end{table}

\begin{table}[!h]
\caption{ 
Parameters of the calculations for $32^3 \times 8$ lattices
}
\label{ensemblesnt8}
\begin{tabular}{ccccccc}
		\hline \hline
$\beta$ &		$T$ (MeV) & $N_{sv,l}$ & $N_{sv,l}^{imp}$ & $N_{sv,s}$ & $N_{sv,s}^{imp}$& \# configurations \\
		\hline
	7.150 &		334.4	& 250	& 0	& 250	& 0	& 4020 \\
	7.280 &		376.8	& 250	& 0 & 250	& 0	& 4080 \\
	7.373 &		410.0	& 250 & 1000	& 250	& 0	& 3940 \\
	7.596 &		500.4	& 250	& 0	& 250	& 0	& 4050 \\
	7.825 &		611.5	& 250	& 0	& 250	& 0	& 3920 \\
	8.000 &		711.3	& 250	& 0	& 250	& 0	& 1800 \\
\hline
\end{tabular}
\end{table}

\begin{table}[!h]
    \caption{
    Parameters of the calculations for $40^3 \times 10$ lattices
    }
    \label{ensemblesnt10}
    \begin{tabular}{ccccccc}
        \hline \hline
        $\beta$ & $T$ (MeV) & $N_{sv,l}$ & $N_{sv,l}^{imp}$ & 
$N_{sv,s}$ & $N_{sv,s}^{imp}$& \# configurations \\
        \hline
        7.373 &    328.0    & 300    & 0    & 300 & 0    & 4060 \\
        7.596 & 400.3    & 300    & 0    & 300 & 0    & 3180 \\
        7.825 & 489.2    & 300    & 0    & 300 & 0    & 3090 \\
        8.000 & 569.1    & 600    & 600    & 200 & 0    & 2670 \\
        8.200 & 675.3    & 200    & 0    & 200 & 0    & 1590 \\
        \hline
    \end{tabular}
\end{table}

\begin{table}[!h]
\caption{ 
Parameters of the calculations for $48^3 \times 12$ lattices
}
\label{ensemblesnt12}
\begin{tabular}{ccccccc}
		\hline \hline
$\beta$ & $T$ (MeV) & $N_{sv,l}$ & $N_{sv,l}^{imp}$ & $N_{sv,s}$ & $N_{sv,s}^{imp}$& \# configurations \\
		\hline
7.596 & 333.6	& 250	& 250	& 250	& 250	& 3360 \\
7.825 & 407.7	& 250	& 250	& 250	& 250	& 3020 \\
8.200 & 562.7	& 250	& 250	& 250	& 250	& 2740 \\
8.400 & 666.9	& 250	& 250	& 250	& 250	& 2710 \\
\hline
\end{tabular}
\end{table}

\newpage

\bibliography{HotQCD}

\begin{thebibliography}{39}%
\makeatletter
\providecommand \@ifxundefined [1]{%
 \@ifx{#1\undefined}
}%
\providecommand \@ifnum [1]{%
 \ifnum #1\expandafter \@firstoftwo
 \else \expandafter \@secondoftwo
 \fi
}%
\providecommand \@ifx [1]{%
 \ifx #1\expandafter \@firstoftwo
 \else \expandafter \@secondoftwo
 \fi
}%
\providecommand \natexlab [1]{#1}%
\providecommand \enquote  [1]{``#1''}%
\providecommand \bibnamefont  [1]{#1}%
\providecommand \bibfnamefont [1]{#1}%
\providecommand \citenamefont [1]{#1}%
\providecommand \href@noop [0]{\@secondoftwo}%
\providecommand \href [0]{\begingroup \@sanitize@url \@href}%
\providecommand \@href[1]{\@@startlink{#1}\@@href}%
\providecommand \@@href[1]{\endgroup#1\@@endlink}%
\providecommand \@sanitize@url [0]{\catcode `\\12\catcode `\$12\catcode
  `\&12\catcode `\#12\catcode `\^12\catcode `\_12\catcode `\%12\relax}%
\providecommand \@@startlink[1]{}%
\providecommand \@@endlink[0]{}%
\providecommand \url  [0]{\begingroup\@sanitize@url \@url }%
\providecommand \@url [1]{\endgroup\@href {#1}{\urlprefix }}%
\providecommand \urlprefix  [0]{URL }%
\providecommand \Eprint [0]{\href }%
\providecommand \doibase [0]{http://dx.doi.org/}%
\providecommand \selectlanguage [0]{\@gobble}%
\providecommand \bibinfo  [0]{\@secondoftwo}%
\providecommand \bibfield  [0]{\@secondoftwo}%
\providecommand \translation [1]{[#1]}%
\providecommand \BibitemOpen [0]{}%
\providecommand \bibitemStop [0]{}%
\providecommand \bibitemNoStop [0]{.\EOS\space}%
\providecommand \EOS [0]{\spacefactor3000\relax}%
\providecommand \BibitemShut  [1]{\csname bibitem#1\endcsname}%
\let\auto@bib@innerbib\@empty
\bibitem [{\citenamefont {Petreczky}(2012)}]{Petreczky:2012rq}%
  \BibitemOpen
  \bibfield  {author} {\bibinfo {author} {\bibfnamefont {P.}~\bibnamefont
  {Petreczky}},\ }\href {\doibase 10.1088/0954-3899/39/9/093002} {\bibfield
  {journal} {\bibinfo  {journal} {J. Phys. G}\ }\textbf {\bibinfo {volume}
  {39}},\ \bibinfo {pages} {093002} (\bibinfo {year} {2012})},\ \Eprint
  {http://arxiv.org/abs/1203.5320} {arXiv:1203.5320 [hep-lat]} \BibitemShut
  {NoStop}%
\bibitem [{\citenamefont {Philipsen}(2013)}]{Philipsen:2012nu}%
  \BibitemOpen
  \bibfield  {author} {\bibinfo {author} {\bibfnamefont {O.}~\bibnamefont
  {Philipsen}},\ }\href {\doibase 10.1016/j.ppnp.2012.09.003} {\bibfield
  {journal} {\bibinfo  {journal} {Prog. Part. Nucl. Phys.}\ }\textbf {\bibinfo
  {volume} {70}},\ \bibinfo {pages} {55} (\bibinfo {year} {2013})},\ \Eprint
  {http://arxiv.org/abs/1207.5999} {arXiv:1207.5999 [hep-lat]} \BibitemShut
  {NoStop}%
\bibitem [{\citenamefont {Ding}\ \emph {et~al.}(2015)\citenamefont {Ding},
  \citenamefont {Karsch},\ and\ \citenamefont {Mukherjee}}]{Ding:2015ona}%
  \BibitemOpen
  \bibfield  {author} {\bibinfo {author} {\bibfnamefont {H.-T.}\ \bibnamefont
  {Ding}}, \bibinfo {author} {\bibfnamefont {F.}~\bibnamefont {Karsch}}, \ and\
  \bibinfo {author} {\bibfnamefont {S.}~\bibnamefont {Mukherjee}},\ }\href@noop
  {} {\  (\bibinfo {year} {2015})},\ \Eprint {http://arxiv.org/abs/1504.05274}
  {arXiv:1504.05274 [hep-lat]} \BibitemShut {NoStop}%
\bibitem [{\citenamefont {Muller}\ and\ \citenamefont
  {Nagle}(2006)}]{Muller:2006ee}%
  \BibitemOpen
  \bibfield  {author} {\bibinfo {author} {\bibfnamefont {B.}~\bibnamefont
  {Muller}}\ and\ \bibinfo {author} {\bibfnamefont {J.~L.}\ \bibnamefont
  {Nagle}},\ }\href {\doibase 10.1146/annurev.nucl.56.080805.140556} {\bibfield
   {journal} {\bibinfo  {journal} {Ann. Rev. Nucl. Part. Sci.}\ }\textbf
  {\bibinfo {volume} {56}},\ \bibinfo {pages} {93} (\bibinfo {year}
  {2006})}\BibitemShut {NoStop}%
\bibitem [{\citenamefont {Linde}(1980)}]{Linde:1980ts}%
  \BibitemOpen
  \bibfield  {author} {\bibinfo {author} {\bibfnamefont {A.~D.}\ \bibnamefont
  {Linde}},\ }\href {\doibase 10.1016/0370-2693(80)90769-8} {\bibfield
  {journal} {\bibinfo  {journal} {Phys. Lett. B}\ }\textbf {\bibinfo {volume}
  {96}},\ \bibinfo {pages} {289} (\bibinfo {year} {1980})}\BibitemShut
  {NoStop}%
\bibitem [{\citenamefont {Gavai}\ and\ \citenamefont
  {Gupta}(2003)}]{Gavai:2002jt}%
  \BibitemOpen
  \bibfield  {author} {\bibinfo {author} {\bibfnamefont {R.~V.}\ \bibnamefont
  {Gavai}}\ and\ \bibinfo {author} {\bibfnamefont {S.}~\bibnamefont {Gupta}},\
  }\href {\doibase 10.1103/PhysRevD.67.034501} {\bibfield  {journal} {\bibinfo
  {journal} {Phys. Rev. D}\ }\textbf {\bibinfo {volume} {67}},\ \bibinfo
  {pages} {034501} (\bibinfo {year} {2003})},\ \Eprint
  {http://arxiv.org/abs/hep-lat/0211015} {arXiv:hep-lat/0211015 [hep-lat]}
  \BibitemShut {NoStop}%
\bibitem [{\citenamefont {Bernard}\ \emph {et~al.}(2005)\citenamefont {Bernard}
  \emph {et~al.}}]{Bernard:2004je}%
  \BibitemOpen
  \bibfield  {author} {\bibinfo {author} {\bibfnamefont {C.}~\bibnamefont
  {Bernard}} \emph {et~al.} (\bibinfo {collaboration} {MILC Collaboration}),\
  }\href {\doibase 10.1103/PhysRevD.71.034504} {\bibfield  {journal} {\bibinfo
  {journal} {Phys. Rev. D}\ }\textbf {\bibinfo {volume} {71}},\ \bibinfo
  {pages} {034504} (\bibinfo {year} {2005})},\ \Eprint
  {http://arxiv.org/abs/hep-lat/0405029} {arXiv:hep-lat/0405029 [hep-lat]}
  \BibitemShut {NoStop}%
\bibitem [{\citenamefont {Allton}\ \emph {et~al.}(2003)\citenamefont {Allton}
  \emph {et~al.}}]{Allton:2003vx}%
  \BibitemOpen
  \bibfield  {author} {\bibinfo {author} {\bibfnamefont {C.}~\bibnamefont
  {Allton}} \emph {et~al.},\ }\href {\doibase 10.1103/PhysRevD.68.014507}
  {\bibfield  {journal} {\bibinfo  {journal} {Phys. Rev. D}\ }\textbf {\bibinfo
  {volume} {68}},\ \bibinfo {pages} {014507} (\bibinfo {year} {2003})},\
  \Eprint {http://arxiv.org/abs/hep-lat/0305007} {arXiv:hep-lat/0305007
  [hep-lat]} \BibitemShut {NoStop}%
\bibitem [{\citenamefont {Allton}\ \emph {et~al.}(2005)\citenamefont {Allton}
  \emph {et~al.}}]{Allton:2005gk}%
  \BibitemOpen
  \bibfield  {author} {\bibinfo {author} {\bibfnamefont {C.}~\bibnamefont
  {Allton}} \emph {et~al.},\ }\href {\doibase 10.1103/PhysRevD.71.054508}
  {\bibfield  {journal} {\bibinfo  {journal} {Phys. Rev. D}\ }\textbf {\bibinfo
  {volume} {71}},\ \bibinfo {pages} {054508} (\bibinfo {year} {2005})},\
  \Eprint {http://arxiv.org/abs/hep-lat/0501030} {arXiv:hep-lat/0501030
  [hep-lat]} \BibitemShut {NoStop}%
\bibitem [{\citenamefont {Cheng}\ \emph {et~al.}(2009)\citenamefont {Cheng}
  \emph {et~al.}}]{Cheng:2008zh}%
  \BibitemOpen
  \bibfield  {author} {\bibinfo {author} {\bibfnamefont {M.}~\bibnamefont
  {Cheng}} \emph {et~al.},\ }\href {\doibase 10.1103/PhysRevD.79.074505}
  {\bibfield  {journal} {\bibinfo  {journal} {Phys. Rev. D}\ }\textbf {\bibinfo
  {volume} {79}},\ \bibinfo {pages} {074505} (\bibinfo {year} {2009})},\
  \Eprint {http://arxiv.org/abs/0811.1006} {arXiv:0811.1006 [hep-lat]}
  \BibitemShut {NoStop}%
\bibitem [{\citenamefont {Bazavov}\ \emph
  {et~al.}(2012{\natexlab{a}})\citenamefont {Bazavov} \emph
  {et~al.}}]{Bazavov:2012vg}%
  \BibitemOpen
  \bibfield  {author} {\bibinfo {author} {\bibfnamefont {A.}~\bibnamefont
  {Bazavov}} \emph {et~al.},\ }\href {\doibase 10.1103/PhysRevLett.109.192302}
  {\bibfield  {journal} {\bibinfo  {journal} {Phys. Rev. Lett.}\ }\textbf
  {\bibinfo {volume} {109}},\ \bibinfo {pages} {192302} (\bibinfo {year}
  {2012}{\natexlab{a}})},\ \Eprint {http://arxiv.org/abs/1208.1220}
  {arXiv:1208.1220 [hep-lat]} \BibitemShut {NoStop}%
\bibitem [{\citenamefont {Bazavov}\ \emph
  {et~al.}(2013{\natexlab{a}})\citenamefont {Bazavov} \emph
  {et~al.}}]{Bazavov:2013dta}%
  \BibitemOpen
  \bibfield  {author} {\bibinfo {author} {\bibfnamefont {A.}~\bibnamefont
  {Bazavov}} \emph {et~al.},\ }\href {\doibase 10.1103/PhysRevLett.111.082301}
  {\bibfield  {journal} {\bibinfo  {journal} {Phys. Rev. Lett.}\ }\textbf
  {\bibinfo {volume} {111}},\ \bibinfo {pages} {082301} (\bibinfo {year}
  {2013}{\natexlab{a}})},\ \Eprint {http://arxiv.org/abs/1304.7220}
  {arXiv:1304.7220 [hep-lat]} \BibitemShut {NoStop}%
\bibitem [{\citenamefont {Bellwied}\ \emph {et~al.}(2013)\citenamefont
  {Bellwied}, \citenamefont {Borsanyi}, \citenamefont {Fodor}, \citenamefont
  {Katz},\ and\ \citenamefont {Ratti}}]{Bellwied:2013cta}%
  \BibitemOpen
  \bibfield  {author} {\bibinfo {author} {\bibfnamefont {R.}~\bibnamefont
  {Bellwied}}, \bibinfo {author} {\bibfnamefont {S.}~\bibnamefont {Borsanyi}},
  \bibinfo {author} {\bibfnamefont {Z.}~\bibnamefont {Fodor}}, \bibinfo
  {author} {\bibfnamefont {S.~D.}\ \bibnamefont {Katz}}, \ and\ \bibinfo
  {author} {\bibfnamefont {C.}~\bibnamefont {Ratti}},\ }\href {\doibase
  10.1103/PhysRevLett.111.202302} {\bibfield  {journal} {\bibinfo  {journal}
  {Phys.Rev.Lett.}\ }\textbf {\bibinfo {volume} {111}},\ \bibinfo {pages}
  {202302} (\bibinfo {year} {2013})},\ \Eprint {http://arxiv.org/abs/1305.6297}
  {arXiv:1305.6297 [hep-lat]} \BibitemShut {NoStop}%
\bibitem [{\citenamefont {Borsanyi}\ \emph {et~al.}(2013)\citenamefont
  {Borsanyi} \emph {et~al.}}]{Borsanyi:2013hza}%
  \BibitemOpen
  \bibfield  {author} {\bibinfo {author} {\bibfnamefont {S.}~\bibnamefont
  {Borsanyi}} \emph {et~al.},\ }\href {\doibase 10.1103/PhysRevLett.111.062005}
  {\bibfield  {journal} {\bibinfo  {journal} {Phys. Rev. Lett.}\ }\textbf
  {\bibinfo {volume} {111}},\ \bibinfo {pages} {062005} (\bibinfo {year}
  {2013})},\ \Eprint {http://arxiv.org/abs/1305.5161} {arXiv:1305.5161
  [hep-lat]} \BibitemShut {NoStop}%
\bibitem [{\citenamefont {Bazavov}\ \emph
  {et~al.}(2014{\natexlab{a}})\citenamefont {Bazavov} \emph
  {et~al.}}]{Bazavov:2014xya}%
  \BibitemOpen
  \bibfield  {author} {\bibinfo {author} {\bibfnamefont {A.}~\bibnamefont
  {Bazavov}} \emph {et~al.},\ }\href {\doibase 10.1103/PhysRevLett.113.072001}
  {\bibfield  {journal} {\bibinfo  {journal} {Phys. Rev. Lett.}\ }\textbf
  {\bibinfo {volume} {113}},\ \bibinfo {pages} {072001} (\bibinfo {year}
  {2014}{\natexlab{a}})},\ \Eprint {http://arxiv.org/abs/1404.6511}
  {arXiv:1404.6511 [hep-lat]} \BibitemShut {NoStop}%
\bibitem [{\citenamefont {Bazavov}\ \emph
  {et~al.}(2014{\natexlab{b}})\citenamefont {Bazavov} \emph
  {et~al.}}]{Bazavov:2014yba}%
  \BibitemOpen
  \bibfield  {author} {\bibinfo {author} {\bibfnamefont {A.}~\bibnamefont
  {Bazavov}} \emph {et~al.},\ }\href {\doibase 10.1016/j.physletb.2014.08.034}
  {\bibfield  {journal} {\bibinfo  {journal} {Phys. Lett. B}\ }\textbf
  {\bibinfo {volume} {737}},\ \bibinfo {pages} {210} (\bibinfo {year}
  {2014}{\natexlab{b}})},\ \Eprint {http://arxiv.org/abs/1404.4043}
  {arXiv:1404.4043 [hep-lat]} \BibitemShut {NoStop}%
\bibitem [{\citenamefont {Gattringer}\ and\ \citenamefont
  {Schadler}(2015)}]{Gattringer:2015hra}%
  \BibitemOpen
  \bibfield  {author} {\bibinfo {author} {\bibfnamefont {C.}~\bibnamefont
  {Gattringer}}\ and\ \bibinfo {author} {\bibfnamefont {H.-P.}\ \bibnamefont
  {Schadler}},\ }\href {\doibase 10.1103/PhysRevD.91.074511} {\bibfield
  {journal} {\bibinfo  {journal} {Phys. Rev. D}\ }\textbf {\bibinfo {volume}
  {91}},\ \bibinfo {pages} {074511} (\bibinfo {year} {2015})},\ \Eprint
  {http://arxiv.org/abs/1411.5133} {arXiv:1411.5133 [hep-lat]} \BibitemShut
  {NoStop}%
\bibitem [{\citenamefont {Blaizot}\ \emph {et~al.}(2001)\citenamefont
  {Blaizot}, \citenamefont {Iancu},\ and\ \citenamefont
  {Rebhan}}]{Blaizot:2001vr}%
  \BibitemOpen
  \bibfield  {author} {\bibinfo {author} {\bibfnamefont {J.}~\bibnamefont
  {Blaizot}}, \bibinfo {author} {\bibfnamefont {E.}~\bibnamefont {Iancu}}, \
  and\ \bibinfo {author} {\bibfnamefont {A.}~\bibnamefont {Rebhan}},\ }\href
  {\doibase 10.1016/S0370-2693(01)01316-8} {\bibfield  {journal} {\bibinfo
  {journal} {Phys. Lett. B}\ }\textbf {\bibinfo {volume} {523}},\ \bibinfo
  {pages} {143} (\bibinfo {year} {2001})},\ \Eprint
  {http://arxiv.org/abs/hep-ph/0110369} {arXiv:hep-ph/0110369 [hep-ph]}
  \BibitemShut {NoStop}%
\bibitem [{\citenamefont {Vuorinen}(2003)}]{Vuorinen:2002ue}%
  \BibitemOpen
  \bibfield  {author} {\bibinfo {author} {\bibfnamefont {A.}~\bibnamefont
  {Vuorinen}},\ }\href {\doibase 10.1103/PhysRevD.67.074032} {\bibfield
  {journal} {\bibinfo  {journal} {Phys. Rev. D}\ }\textbf {\bibinfo {volume}
  {67}},\ \bibinfo {pages} {074032} (\bibinfo {year} {2003})},\ \Eprint
  {http://arxiv.org/abs/hep-ph/0212283} {arXiv:hep-ph/0212283 [hep-ph]}
  \BibitemShut {NoStop}%
\bibitem [{\citenamefont {Rebhan}(2003)}]{Rebhan:2003fj}%
  \BibitemOpen
  \bibfield  {author} {\bibinfo {author} {\bibfnamefont {A.}~\bibnamefont
  {Rebhan}},\ }\href@noop {} {\emph {\bibinfo {title} {Strong and Electroweak
  Matter 2002}}}\ (\bibinfo  {publisher} {World Scientific},\ \bibinfo {year}
  {2003})\ \Eprint {http://arxiv.org/abs/hep-ph/0301130} {arXiv:hep-ph/0301130
  [hep-ph]} \BibitemShut {NoStop}%
\bibitem [{\citenamefont {Hietanen}\ and\ \citenamefont
  {Rummukainen}(2008)}]{Hietanen:2008xb}%
  \BibitemOpen
  \bibfield  {author} {\bibinfo {author} {\bibfnamefont {A.}~\bibnamefont
  {Hietanen}}\ and\ \bibinfo {author} {\bibfnamefont {K.}~\bibnamefont
  {Rummukainen}},\ }\href {\doibase 10.1088/1126-6708/2008/04/078} {\bibfield
  {journal} {\bibinfo  {journal} {J. High Energy Phys.}\ }\textbf {\bibinfo
  {volume} {04}},\ \bibinfo {pages} {078} (\bibinfo {year} {2008})},\ \Eprint
  {http://arxiv.org/abs/0802.3979} {arXiv:0802.3979 [hep-lat]} \BibitemShut
  {NoStop}%
\bibitem [{\citenamefont {Haque}\ and\ \citenamefont
  {Mustafa}()}]{Haque:2010rb}%
  \BibitemOpen
  \bibfield  {author} {\bibinfo {author} {\bibfnamefont {N.}~\bibnamefont
  {Haque}}\ and\ \bibinfo {author} {\bibfnamefont {M.~G.}\ \bibnamefont
  {Mustafa}},\ }\href@noop {} {\ }\Eprint {http://arxiv.org/abs/1007.2076}
  {arXiv:1007.2076 [hep-ph]} \BibitemShut {NoStop}%
\bibitem [{\citenamefont {Andersen}\ \emph {et~al.}(2013)\citenamefont
  {Andersen}, \citenamefont {Mogliacci}, \citenamefont {Su},\ and\
  \citenamefont {Vuorinen}}]{Andersen:2012wr}%
  \BibitemOpen
  \bibfield  {author} {\bibinfo {author} {\bibfnamefont {J.~O.}\ \bibnamefont
  {Andersen}}, \bibinfo {author} {\bibfnamefont {S.}~\bibnamefont {Mogliacci}},
  \bibinfo {author} {\bibfnamefont {N.}~\bibnamefont {Su}}, \ and\ \bibinfo
  {author} {\bibfnamefont {A.}~\bibnamefont {Vuorinen}},\ }\href {\doibase
  10.1103/PhysRevD.87.074003} {\bibfield  {journal} {\bibinfo  {journal} {Phys.
  Rev. D}\ }\textbf {\bibinfo {volume} {87}},\ \bibinfo {pages} {074003}
  (\bibinfo {year} {2013})},\ \Eprint {http://arxiv.org/abs/1210.0912}
  {arXiv:1210.0912 [hep-ph]} \BibitemShut {NoStop}%
\bibitem [{\citenamefont {Mogliacci}\ \emph {et~al.}(2013)\citenamefont
  {Mogliacci}, \citenamefont {Andersen}, \citenamefont {Strickland},
  \citenamefont {Su},\ and\ \citenamefont {Vuorinen}}]{Mogliacci:2013mca}%
  \BibitemOpen
  \bibfield  {author} {\bibinfo {author} {\bibfnamefont {S.}~\bibnamefont
  {Mogliacci}}, \bibinfo {author} {\bibfnamefont {J.~O.}\ \bibnamefont
  {Andersen}}, \bibinfo {author} {\bibfnamefont {M.}~\bibnamefont
  {Strickland}}, \bibinfo {author} {\bibfnamefont {N.}~\bibnamefont {Su}}, \
  and\ \bibinfo {author} {\bibfnamefont {A.}~\bibnamefont {Vuorinen}},\ }\href
  {\doibase 10.1007/Journal of High Energy Phys.12(2013)055} {\bibfield
  {journal} {\bibinfo  {journal} {Journal of High Energy Phys.}\ }\textbf
  {\bibinfo {volume} {12}},\ \bibinfo {pages} {055} (\bibinfo {year} {2013})},\
  \Eprint {http://arxiv.org/abs/1307.8098} {arXiv:1307.8098 [hep-ph]}
  \BibitemShut {NoStop}%
\bibitem [{\citenamefont {Haque}\ \emph {et~al.}(2013)\citenamefont {Haque},
  \citenamefont {Mustafa},\ and\ \citenamefont {Strickland}}]{Haque:2013qta}%
  \BibitemOpen
  \bibfield  {author} {\bibinfo {author} {\bibfnamefont {N.}~\bibnamefont
  {Haque}}, \bibinfo {author} {\bibfnamefont {M.~G.}\ \bibnamefont {Mustafa}},
  \ and\ \bibinfo {author} {\bibfnamefont {M.}~\bibnamefont {Strickland}},\
  }\href {\doibase 10.1007/Journal of High Energy Phys.07(2013)184} {\bibfield
  {journal} {\bibinfo  {journal} {Journal of High Energy Phys.}\ }\textbf
  {\bibinfo {volume} {07}},\ \bibinfo {pages} {184} (\bibinfo {year} {2013})},\
  \Eprint {http://arxiv.org/abs/1302.3228} {arXiv:1302.3228 [hep-ph]}
  \BibitemShut {NoStop}%
\bibitem [{\citenamefont {Haque}\ \emph
  {et~al.}(2014{\natexlab{a}})\citenamefont {Haque}, \citenamefont {Andersen},
  \citenamefont {Mustafa}, \citenamefont {Strickland},\ and\ \citenamefont
  {Su}}]{Haque:2013sja}%
  \BibitemOpen
  \bibfield  {author} {\bibinfo {author} {\bibfnamefont {N.}~\bibnamefont
  {Haque}}, \bibinfo {author} {\bibfnamefont {J.~O.}\ \bibnamefont {Andersen}},
  \bibinfo {author} {\bibfnamefont {M.~G.}\ \bibnamefont {Mustafa}}, \bibinfo
  {author} {\bibfnamefont {M.}~\bibnamefont {Strickland}}, \ and\ \bibinfo
  {author} {\bibfnamefont {N.}~\bibnamefont {Su}},\ }\href {\doibase
  10.1103/PhysRevD.89.061701} {\bibfield  {journal} {\bibinfo  {journal} {Phys.
  Rev. D}\ }\textbf {\bibinfo {volume} {89}},\ \bibinfo {pages} {061701}
  (\bibinfo {year} {2014}{\natexlab{a}})},\ \Eprint
  {http://arxiv.org/abs/1309.3968} {arXiv:1309.3968 [hep-ph]} \BibitemShut
  {NoStop}%
\bibitem [{\citenamefont {Haque}\ \emph
  {et~al.}(2014{\natexlab{b}})\citenamefont {Haque}, \citenamefont
  {Bandyopadhyay}, \citenamefont {Andersen}, \citenamefont {Mustafa},
  \citenamefont {Strickland} \emph {et~al.}}]{Haque:2014rua}%
  \BibitemOpen
  \bibfield  {author} {\bibinfo {author} {\bibfnamefont {N.}~\bibnamefont
  {Haque}}, \bibinfo {author} {\bibfnamefont {A.}~\bibnamefont
  {Bandyopadhyay}}, \bibinfo {author} {\bibfnamefont {J.~O.}\ \bibnamefont
  {Andersen}}, \bibinfo {author} {\bibfnamefont {M.~G.}\ \bibnamefont
  {Mustafa}}, \bibinfo {author} {\bibfnamefont {M.}~\bibnamefont {Strickland}},
   \emph {et~al.},\ }\href {\doibase 10.1007/Journal of High Energy
  Phys.05(2014)027} {\bibfield  {journal} {\bibinfo  {journal} {Journal of High
  Energy Phys.}\ }\textbf {\bibinfo {volume} {05}},\ \bibinfo {pages} {027}
  (\bibinfo {year} {2014}{\natexlab{b}})},\ \Eprint
  {http://arxiv.org/abs/1402.6907} {arXiv:1402.6907 [hep-ph]} \BibitemShut
  {NoStop}%
\bibitem [{\citenamefont {Bazavov}\ \emph
  {et~al.}(2013{\natexlab{b}})\citenamefont {Bazavov} \emph
  {et~al.}}]{Bazavov:2013uja}%
  \BibitemOpen
  \bibfield  {author} {\bibinfo {author} {\bibfnamefont {A.}~\bibnamefont
  {Bazavov}} \emph {et~al.},\ }\href {\doibase 10.1103/PhysRevD.88.094021}
  {\bibfield  {journal} {\bibinfo  {journal} {Phys.Rev. D}\ }\textbf {\bibinfo
  {volume} {88}},\ \bibinfo {pages} {094021} (\bibinfo {year}
  {2013}{\natexlab{b}})},\ \Eprint {http://arxiv.org/abs/1309.2317}
  {arXiv:1309.2317 [hep-lat]} \BibitemShut {NoStop}%
\bibitem [{\citenamefont {Follana}\ \emph {et~al.}(2007)\citenamefont {Follana}
  \emph {et~al.}}]{Follana:2006rc}%
  \BibitemOpen
  \bibfield  {author} {\bibinfo {author} {\bibfnamefont {E.}~\bibnamefont
  {Follana}} \emph {et~al.} (\bibinfo {collaboration} {HPQCD Collaboration,
  UKQCD Collaboration}),\ }\href {\doibase 10.1103/PhysRevD.75.054502}
  {\bibfield  {journal} {\bibinfo  {journal} {Phys. Rev. D}\ }\textbf {\bibinfo
  {volume} {75}},\ \bibinfo {pages} {054502} (\bibinfo {year} {2007})},\
  \Eprint {http://arxiv.org/abs/hep-lat/0610092} {arXiv:hep-lat/0610092
  [hep-lat]} \BibitemShut {NoStop}%
\bibitem [{\citenamefont {Bazavov}\ \emph
  {et~al.}(2012{\natexlab{b}})\citenamefont {Bazavov} \emph
  {et~al.}}]{Bazavov:2011nk}%
  \BibitemOpen
  \bibfield  {author} {\bibinfo {author} {\bibfnamefont {A.}~\bibnamefont
  {Bazavov}} \emph {et~al.},\ }\href {\doibase 10.1103/PhysRevD.85.054503}
  {\bibfield  {journal} {\bibinfo  {journal} {Phys. Rev. D}\ }\textbf {\bibinfo
  {volume} {85}},\ \bibinfo {pages} {054503} (\bibinfo {year}
  {2012}{\natexlab{b}})},\ \Eprint {http://arxiv.org/abs/1111.1710}
  {arXiv:1111.1710 [hep-lat]} \BibitemShut {NoStop}%
\bibitem [{\citenamefont {Bazavov}\ \emph
  {et~al.}(2014{\natexlab{c}})\citenamefont {Bazavov} \emph
  {et~al.}}]{Bazavov:2014pvz}%
  \BibitemOpen
  \bibfield  {author} {\bibinfo {author} {\bibfnamefont {A.}~\bibnamefont
  {Bazavov}} \emph {et~al.} (\bibinfo {collaboration} {HotQCD}),\ }\href
  {\doibase 10.1103/PhysRevD.90.094503} {\bibfield  {journal} {\bibinfo
  {journal} {Phys.Rev.}\ }\textbf {\bibinfo {volume} {D90}},\ \bibinfo {pages}
  {094503} (\bibinfo {year} {2014}{\natexlab{c}})},\ \Eprint
  {http://arxiv.org/abs/1407.6387} {arXiv:1407.6387 [hep-lat]} \BibitemShut
  {NoStop}%
\bibitem [{\citenamefont {Clark}\ \emph {et~al.}(2005)\citenamefont {Clark},
  \citenamefont {Kennedy},\ and\ \citenamefont {Sroczynski}}]{Clark:2004cp}%
  \BibitemOpen
  \bibfield  {author} {\bibinfo {author} {\bibfnamefont {M.}~\bibnamefont
  {Clark}}, \bibinfo {author} {\bibfnamefont {A.}~\bibnamefont {Kennedy}}, \
  and\ \bibinfo {author} {\bibfnamefont {Z.}~\bibnamefont {Sroczynski}},\
  }\href {\doibase 10.1016/j.nuclphysbps.2004.11.192} {\bibfield  {journal}
  {\bibinfo  {journal} {Nucl. Phys. Proc. Suppl.}\ }\textbf {\bibinfo {volume}
  {140}},\ \bibinfo {pages} {835} (\bibinfo {year} {2005})},\ \Eprint
  {http://arxiv.org/abs/hep-lat/0409133} {arXiv:hep-lat/0409133 [hep-lat]}
  \BibitemShut {NoStop}%
\bibitem [{\citenamefont {Allton}\ \emph {et~al.}(2002)\citenamefont {Allton}
  \emph {et~al.}}]{Allton:2002zi}%
  \BibitemOpen
  \bibfield  {author} {\bibinfo {author} {\bibfnamefont {C.}~\bibnamefont
  {Allton}} \emph {et~al.},\ }\href {\doibase 10.1103/PhysRevD.66.074507}
  {\bibfield  {journal} {\bibinfo  {journal} {Phys. Rev. D}\ }\textbf {\bibinfo
  {volume} {66}},\ \bibinfo {pages} {074507} (\bibinfo {year} {2002})},\
  \Eprint {http://arxiv.org/abs/hep-lat/0204010} {arXiv:hep-lat/0204010
  [hep-lat]} \BibitemShut {NoStop}%
\bibitem [{\citenamefont {http://www.r project.org/}()}]{Rpackage}%
  \BibitemOpen
  \bibfield  {author} {\bibinfo {author} {\bibnamefont {http://www.r
  project.org/}},\ }\href@noop {} {\ }\BibitemShut {NoStop}%
\bibitem [{\citenamefont {Hegde}\ \emph {et~al.}(2008)\citenamefont {Hegde},
  \citenamefont {Karsch}, \citenamefont {Laermann},\ and\ \citenamefont
  {Shcheredin}}]{Hegde:2008nx}%
  \BibitemOpen
  \bibfield  {author} {\bibinfo {author} {\bibfnamefont {P.}~\bibnamefont
  {Hegde}}, \bibinfo {author} {\bibfnamefont {F.}~\bibnamefont {Karsch}},
  \bibinfo {author} {\bibfnamefont {E.}~\bibnamefont {Laermann}}, \ and\
  \bibinfo {author} {\bibfnamefont {S.}~\bibnamefont {Shcheredin}},\ }\href
  {\doibase 10.1140/epjc/s10052-008-0613-7} {\bibfield  {journal} {\bibinfo
  {journal} {Eur. Phys. J.}\ }\textbf {\bibinfo {volume} {C55}},\ \bibinfo
  {pages} {423} (\bibinfo {year} {2008})},\ \Eprint
  {http://arxiv.org/abs/0801.4883} {arXiv:0801.4883 [hep-lat]} \BibitemShut
  {NoStop}%
\bibitem [{\citenamefont {Braaten}\ and\ \citenamefont
  {Nieto}(1996)}]{Braaten:1995jr}%
  \BibitemOpen
  \bibfield  {author} {\bibinfo {author} {\bibfnamefont {E.}~\bibnamefont
  {Braaten}}\ and\ \bibinfo {author} {\bibfnamefont {A.}~\bibnamefont
  {Nieto}},\ }\href {\doibase 10.1103/PhysRevD.53.3421} {\bibfield  {journal}
  {\bibinfo  {journal} {Phys. Rev. D}\ }\textbf {\bibinfo {volume} {53}},\
  \bibinfo {pages} {3421} (\bibinfo {year} {1996})},\ \Eprint
  {http://arxiv.org/abs/hep-ph/9510408} {arXiv:hep-ph/9510408 [hep-ph]}
  \BibitemShut {NoStop}%
\bibitem [{\citenamefont {Kajantie}\ \emph {et~al.}(1997)\citenamefont
  {Kajantie}, \citenamefont {Laine}, \citenamefont {Rummukainen},\ and\
  \citenamefont {Shaposhnikov}}]{Kajantie:1997tt}%
  \BibitemOpen
  \bibfield  {author} {\bibinfo {author} {\bibfnamefont {K.}~\bibnamefont
  {Kajantie}}, \bibinfo {author} {\bibfnamefont {M.}~\bibnamefont {Laine}},
  \bibinfo {author} {\bibfnamefont {K.}~\bibnamefont {Rummukainen}}, \ and\
  \bibinfo {author} {\bibfnamefont {M.~E.}\ \bibnamefont {Shaposhnikov}},\
  }\href {\doibase 10.1016/S0550-3213(97)00425-2} {\bibfield  {journal}
  {\bibinfo  {journal} {Nucl. Phys. B}\ }\textbf {\bibinfo {volume} {503}},\
  \bibinfo {pages} {357} (\bibinfo {year} {1997})},\ \Eprint
  {http://arxiv.org/abs/hep-ph/9704416} {arXiv:hep-ph/9704416 [hep-ph]}
  \BibitemShut {NoStop}%
\bibitem [{\citenamefont {Laine}\ and\ \citenamefont
  {Schroder}(2005)}]{Laine:2005ai}%
  \BibitemOpen
  \bibfield  {author} {\bibinfo {author} {\bibfnamefont {M.}~\bibnamefont
  {Laine}}\ and\ \bibinfo {author} {\bibfnamefont {Y.}~\bibnamefont
  {Schroder}},\ }\href {\doibase 10.1088/1126-6708/2005/03/067} {\bibfield
  {journal} {\bibinfo  {journal} {Journal of High Energy Phys.}\ }\textbf
  {\bibinfo {volume} {03}},\ \bibinfo {pages} {067} (\bibinfo {year} {2005})},\
  \Eprint {http://arxiv.org/abs/hep-ph/0503061} {arXiv:hep-ph/0503061 [hep-ph]}
  \BibitemShut {NoStop}%
\bibitem [{\citenamefont {Bellwied}\ \emph {et~al.}(2015)\citenamefont
  {Bellwied}, \citenamefont {Borsanyi}, \citenamefont {Fodor}, \citenamefont
  {Katz}, \citenamefont {Pasztor}, \citenamefont {Ratti},\ and\ \citenamefont
  {Szabo}}]{Bellwied:2015lba}%
  \BibitemOpen
  \bibfield  {author} {\bibinfo {author} {\bibfnamefont {R.}~\bibnamefont
  {Bellwied}}, \bibinfo {author} {\bibfnamefont {S.}~\bibnamefont {Borsanyi}},
  \bibinfo {author} {\bibfnamefont {Z.}~\bibnamefont {Fodor}}, \bibinfo
  {author} {\bibfnamefont {S.~D.}\ \bibnamefont {Katz}}, \bibinfo {author}
  {\bibfnamefont {A.}~\bibnamefont {Pasztor}}, \bibinfo {author} {\bibfnamefont
  {C.}~\bibnamefont {Ratti}}, \ and\ \bibinfo {author} {\bibfnamefont {K.~K.}\
  \bibnamefont {Szabo}},\ }\href@noop {} {\  (\bibinfo {year} {2015})},\
  \Eprint {http://arxiv.org/abs/1507.04627} {arXiv:1507.04627 [hep-lat]}
  \BibitemShut {NoStop}%
\end{thebibliography}%

\end{document}